# Deformation and necking of liquid droplets in a magnetic field.

Sruthy Poulose, Jennifer A. Quirke, Plamen Stamenov, Matthias E. Möbius and J. M. D. Coey*

School of Physics, Trinity College, Dublin 2, Ireland.

**Abstract.**

Pendant droplets of water and paramagnetic solutions are studied in the presence of uniform and nonuniform magnetic fields produced by small permanent magnet arrays, both in static conditions and during dynamic pinch-off. Static measurements of the droplet shape are analysed in terms of an apparent surface tension $\gamma_{app}$ or an effective density $\rho_{eff}$. The change of surface tension of deionized water in a uniform field of 450 mT is insignificant, $0.19 \pm 0.21$ mNm$^{-1}$. Measurements on droplets of compensated zero-susceptibility solutions of Cu$^{2+}$, Mn$^{2+}$ and Dy$^{3+}$ where the shape is unaffected by any magnetic body force show changes of surface tension of about -1% in 500 mT. Magnetic field gradients of up to 100 T$^2$m$^{-1}$ deform the droplets and lead to changes of $\rho_{eff}$ that are negative for diamagnetic solutions (buoyancy effect) and positive for paramagnetic solutions. The droplet profile of strongly-paramagnetic 0.1 Dy M DyCl$_3$ solution is analysed, treating the nonuniform vertical field gradient as a spatial variation of gravity. The influence of Maxwell stress on droplet shape is discussed. In dynamic measurements, the droplet shape at pinch-off is recorded by high-speed photography and analysed in terms of a relative change of dynamic surface tension in the presence of a magnetic field. The surface-tension-dependent pre-factor of the scaling law that governs the pinch-off dynamics shows no difference for pure water or 0.11 M DyCl$_3$ solutions in the field. The nonuniform field has no influence in the pinch-off region because the filament diameter is much less than the capillary length.

*Corresponding author, jcoey@tcd.ie

# 1. Introduction

The surface tension of liquids plays a crucial role in wetting and interfacial phenomena. Microscale and nanoscale device applications such as micro-mechatronics, fluidic micro-robotics and precision micro-manipulators all depend on the interfacial or surface tension of water-based fluids [1]. The functioning of the human body itself depends critically on the surface tension of different biological fluids and surfactants. The flow properties of liquids can be manipulated by external stimuli such as pH [2], light [3], surface tension or wettability gradients [4] and external magnetic [5,6] or electric fields [6,7]. It is important to know if and how magnetic fields can influence the surface tension of liquids and, more generally, influence the shape of liquid droplets on which common methods for measuring surface tension depend.

The existence of any effect of a magnetic field on the surface tension of water is a controversial topic that has been addressed by many researchers over the years. There is no consensus [8-11], and an adequate explanation is still missing [10,12-14]. Fujimora et al [10], for example, reported an increase of 1.32 mN/m for the surface tension of water in a field of 10 T using a surface-wave resonance method [10]. Others found a decrease in surface tension of Millipore water with magnetic field exposure using a commercial magnetic water conditioner [8], surface wave resonator[10] and neodymium ring magnets[9]. One report states that the magnetic field effects depend on the circulation time and the effect increases with field duration[13]. However, in a recent report Hayakawa et al applied a uniform field of 250 mT to pendant droplets using permanent magnets [12] and found that the surface tension of water was essentially unaffected, but there was a small negative effect in a 0.1 M Ho solution. There are numerical studies of pendant droplet shape in uniform and nonuniform magnetic fields using a modified Young-Laplace equation [15]. Studies conducted by Katsuki *et al* [16] on magnetic field effects in pseudo-microgravity conditions using a superconducting magnet with a field of 15 T established that the size and shape of the drop changes due to the high magnetic field gradient forces.

     Here, we extend the scope of the investigations of effects of magnetic fields on the shape of pendant droplets and the static surface tension of water to include several paramagnetic aqueous solutions, and also neck shape and dynamic surface tension during droplet detachment. The static surface tension is the thermodynamic property of a liquid that governs its equilibrium shape. Water, however, was thought to exhibit a different, dynamic surface tension [17] during the pinch-off process as a pendant droplet becomes detached and falls from an orifice. When a new surfactant

interface is formed, its initial surface tension will decrease as surfactant molecules are transported to the interface until the equilibrium value is re-established [18]. During the pinch-off of a pure liquid, the dynamic surface tension inferred is the value at an interface that is in the process of forming the fluid filament, which occurs on a millisecond timescale. No discrepancy between the static and dynamic surface tension was expected for water or other pure liquids, but Hauner's results[17] suggested a higher dynamic value for water of ∼ 90 mNm$^{-1}$, compared to the static value of 72.8 mNm$^{-1}$. While further studies cast doubt on the explicit values measured by the pinch-off method [19], it can be used to investigate any relative differences in surface tension due to the magnetic field on time scales accessible by high-speed photography. In our study, we have explored the magnetic field effects on liquid droplets in the micro and macroscale regimes, and discuss the results in terms of magnetic field gradient forces and static and dynamic surface tension.

## 2. Experimental Methods

### 2.1 Static Measurements

Pendant droplets were measured in a commercial DataPhysics OCA 25 analyser that records the shape of a pear-shaped pendant droplet (Fig. 1), which is assumed to depend only on the *density* ρ of the liquid and its *surface tension* γ. The Young - Laplace pressure difference at a point P on the droplet surface with principal radii of curvature $R_1$ and $R_2$ is

$$\Delta P = \gamma(1/R_1 + 1/R_2) \quad (1)$$

and there is an additional vertical pressure gradient $dP/dz = \rho g$ due to gravity. If the lowest point of the droplet is O, the arc length OP is *s* and the tangent to the droplet in a vertical plane containing O and P makes an angle Φ with the horizontal, the surface tension is obtained by numerically fitting an image of the droplet shape to the three parametric equations (2-4), a procedure carried out automatically by software supplied by the manufacturer.

$$\frac{d\Phi}{ds} = -\frac{\sin\Phi}{x} + \frac{2}{R} + \frac{\Delta\rho g z}{\gamma} \quad (2)$$

$$\frac{dx}{ds} = \cos\Phi \quad (3)$$

$$\frac{dz}{ds} = \sin \phi \qquad (4)$$

where $2/R = (1/R_1 + 1/R_2)$

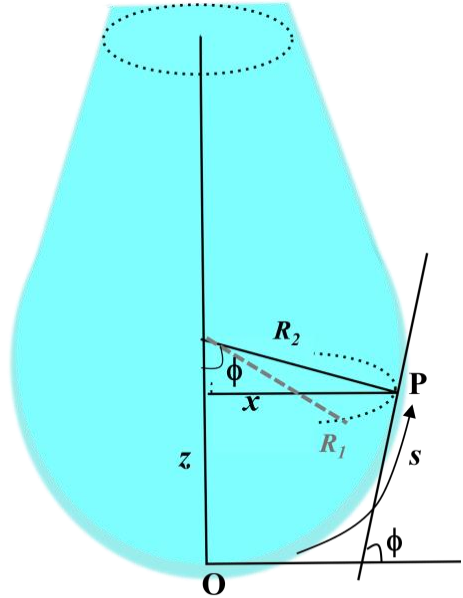

Figure 1. A pendant droplet.

Surface tension tries to minimise the surface area of a drop and makes it spherical, whereas gravity stretches the drop and makes it pear-shaped. Given a value of $\rho$, the analyser will compute $\gamma$ from the droplet shape. We use a nonmagnetic quartz nozzle with internal diameter 1.75 mm and outer diameter 3.02 mm. The lower surface is polished. The volume of liquid in the droplet is approximately 50 µL. Our procedure is to input the density of the liquid to determine the surface tension in zero field $\gamma_0$, and then repeat the measurement in some configuration of magnetic field **B**, noting the difference $\Delta\gamma_{app}$ given by the analyser. The quantity $\Delta\gamma_{app}$ is therefore a shape parameter, not necessarily due to a real change of surface tension.

The introduction of inhomogeneous field gradients to the problem results in a spatial variation of the body force experienced by the drop. This can be explicitly taken into account in the modelling by adding a $\Delta\rho$ term to the right-hand side of Eq. 2 of the form, $[g + (dg/dz).z](z/\gamma)$ that takes into account, not any changes of the gravitational acceleration, but rather changes of the magnetic field gradient produced along the main, vertical $z$-axis. We have included this additional term in a customized version of the differential solver used to integrate Eq. 1- 4, within a MathCad$^{TM}$ computational and fitting routine. The ability to extract this additional parameter $\delta = dg/dz$, depends on the level of pixel noise in the images acquired of the drops and the overall non-trivial droplet shape deformation, which is produced by the gradient-field magnet assemblies used.

In view of the small size of the droplet and the limited space available, we used configurations of permanent magnets to generate uniform or nonuniform fields of up to 500 mT. Unlike small electromagnets, permanent magnet arrays generate no heat and cause no convection. The three configurations illustrated in Fig. 3 are all made of Nd-Fe-B with a remanence of 1.26 T. a) is a uniform horizontal flux source with two $50 \times 20 \times 10$ mm³ magnets and permalloy pole pieces producing a field of up to 450 mT in the *y*-direction, depending on the magnet separation. b) is an array of five 10 mm magnet cubes that produces a vertical field $B_z$ and a negative vertical field gradient $dB_z/dz$ along the z-axis. c) is an array of three 10 mm magnet cubes that produces a horizontal field $B_x$ and a negative vertical field gradient $dB_x/dz$ along the z-axis.

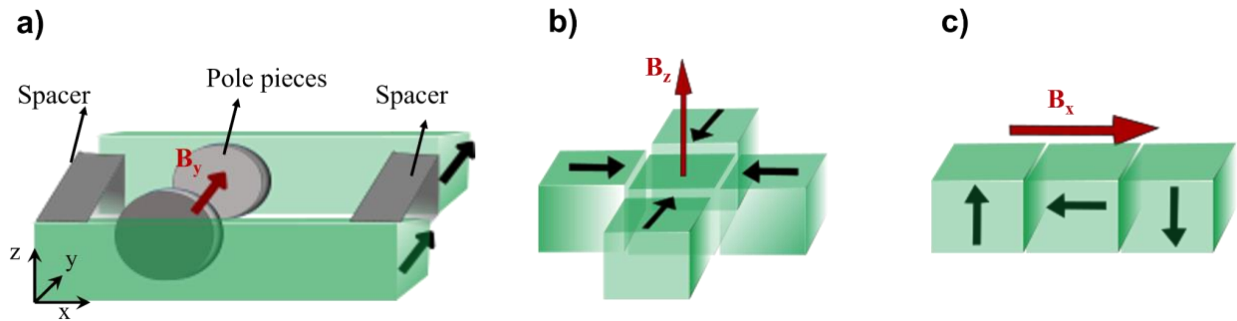

Figure 2. Permanent magnet arrays used to apply magnetic fields to the pendant drop. a) Uniform horizontal field, b) Vertical field with a vertical gradient c) Horizontal field with a vertical gradient. Black arrows indicate the directions of magnetization of the magnet blocks

Fig. 3 shows plots of $B$ and $dB/dz$ calculated as a function of distance from the surface of the 5-magnet and 3-magnet arrays. The magnet blocks were represented by uniform sheers of magnetic charge on the surfaces perpendicular to the magnetization $\boldsymbol{M}$. Three easurement positions of the centre of the droplet are indicated on the curves by $V_{1,2,3}$ for the 5-magnet array and $H_{1,2,3}$ for the 3-magnet array.

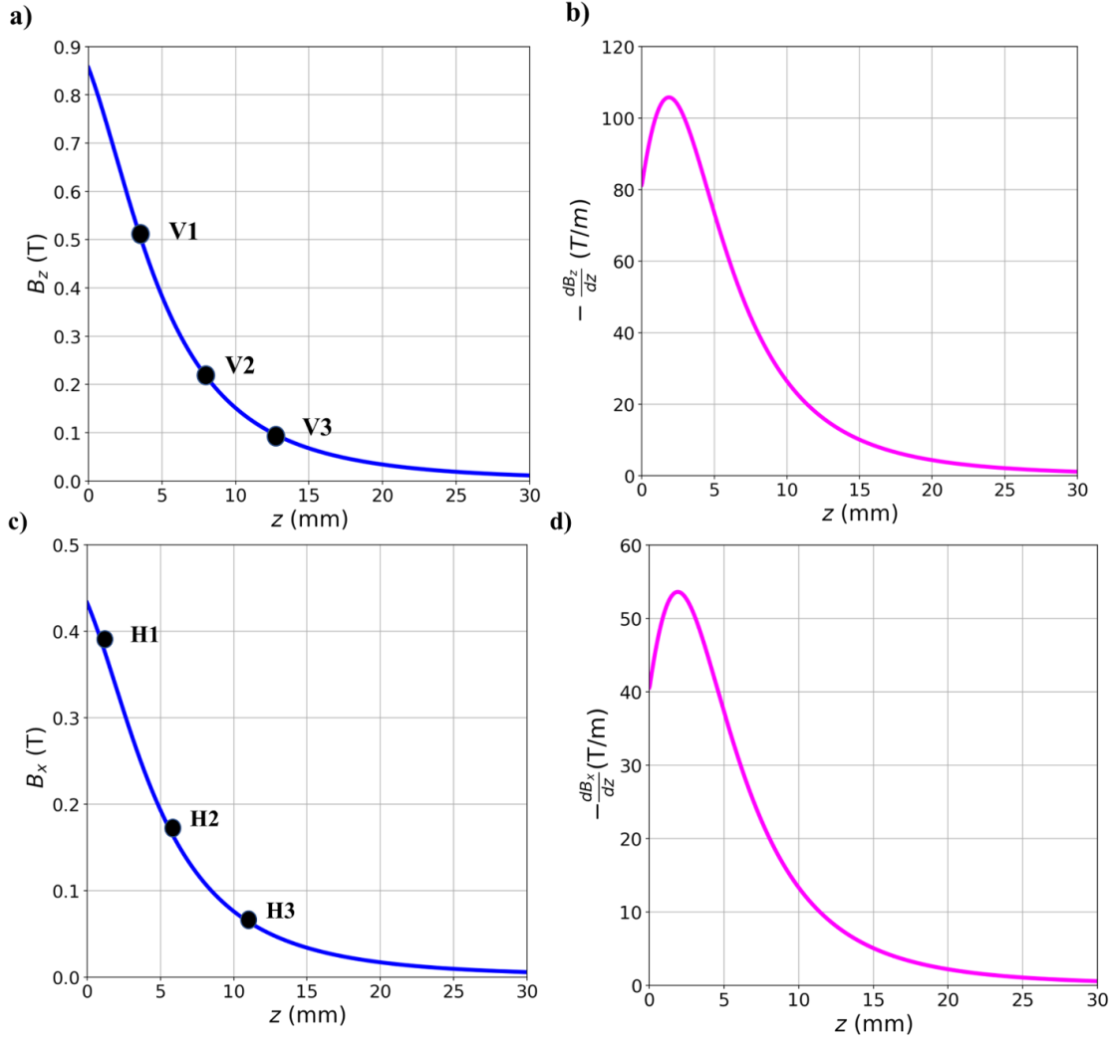

Figure 3. Field plots for the 5-magnet array, with the values of a) $B_z$ and b) -$dB_z/dz$ along the vertical axis and the 3-magnet array, with the values of c) $B_x$ and d) -$dB_z/dz$ along the vertical axis.

When a magnetic field is present, there are three effects that could influence the shape in addition to surface tension and gravity. They are:

i) A change in surface tension of the liquid induced by the magnetic field
ii) The Kelvin force $\mathbf{F}_K = \chi \nabla B^2 / 2\mu_0$,
iii) Maxwell stress

where $\chi$ is the dimensionless SI susceptibility of the liquid ($\chi_w$ = -9 x 10$^{-6}$ for water) and $\mu_0$ is the magnetic constant. For all the liquids we are concerned with here, $\chi \ll 1$ and any effects on surface tension are linear in $B$.

The Kelvin force (also known as the magnetic field gradient force or $B\nabla B$ force) is therefore vertical for the arrays in Fig. 2b) and c) and the gradient is negative; its effect is to reduce the effective density $\rho_{eff}$ of diamagnetic liquids and increase it for paramagnetic liquids. When $\nabla_z B^2$ is big enough, it is possible to levitate water completely; the effective change of density $\Delta\rho_{eff}$ is given by a force-balance equation, where g is the acceleration due to gravity.

$$\Delta\rho_{eff} g = \chi B \nabla_z B / \mu_0 \qquad (5)$$

If, for example, $B\nabla_z B$ = 100 T$^2$m$^{-1}$, the apparent density of water is reduced by 7%. The levitation condition for water $\Delta\rho_{eff} = -\rho$ is $B\nabla_z B$ = 1360 T$^2$m$^{-1}$.

In order to to investigate the magnetic field effects on surface tension in paramagnetic solutions, we have adopted a zero-susceptibility method, where we make use of the fact that the dimensionless susceptibility of water is negative, $\chi_w$ = -9 x 10$^{-6}$, whereas that of 3$d$ or 4$f$ ions is positive and follows a Curie law

$$\chi = \mu_0 n g^2 p_{eff}^2 \mu_B^2 / 3 k_B T. \qquad (6)$$

Here n is the number of ions per unit volume, g is the Landé g-factor and the effective Bohr magneton number p$_{eff}$ is $(S(S+1))^{1/2}$ for 3$d$ ions and $(J(J+1))^{1/2}$ for 4$f$ ions where $S$ and $J$ are the spin quantum number and the total angular momentum quantum number, respectively [20]. The effective ionic moment is p$_{eff}\mu_B$ where $\mu_B$ is the Bohr magneton. The molar susceptibility $\chi_{mol}$ is obtained by replacing n by Avogadro's number N$_0$. The numerical expression for the susceptibility of a mole of ions is $\chi_{mol}$ = 1.571x10$^{-6}$ p$_{eff}^2$/T. The susceptibility of an aqueous ionic solution of molarity $x$ in the dilute limit is therefore

$$\chi = \chi_w/1000 + 1.571 \; 10^{-6} \; x p_{eff}^2 / T. \qquad (7)$$

The first term is the susceptibility of a litre of water. It follows that there is a molar concentration x$_0$ of any paramagnetic ion where the net susceptibility is zero. In this case, Maxwell stress is absent and there are no Kelvin force effects. Any change in the shape of the droplet must be attributed to a change in the surface tension of the liquid. Table 1 lists the molar susceptibility of the three ions we will consider, $Cu^{2+}$, $Mn^{2+}$ and $Dy^{3+}$, as well as the molarity $x_0 = -9 \times 10^{-9}/\chi_{mol}$ of the zero-susceptibility solution where $\chi_{mol}$ is the molar susceptibility of the ions, given by Eq. 6 with $n = N_0$ and using a Landé g-factor of 2 for $3d$ ions and 4/3 for $Dy^{3+}$. Results for the three ions at $T = 295$ K are shown in Table I. Also included are the experimentally measured values $x_0^{exp}$, which are discussed in §3.1.

Table I Calculated susceptibility at 295 K, and calculated and measured zero-susceptibility concentrations

|  | $Cu^{2+}$ | $Mn^{2+}$ | $Dy^{3+}$ |
| --- | --- | --- | --- |
| $m_{eff}$ | 1.73 | 5.92 | 10.65 |
| $\chi_{mol}$ | $16 \times 10^{-9}$ | $187 \times 10^{-9}$ | $604 \times 10^{-9}$ |
| $x_0$ | 0.56 | 0.048 | 0.015 |
| $x_0^{exp}$ | 0.47 | 0.051 | 0.017 |

## 2.2 Dynamic measurements

Dynamic surface tension was probed by recording the pinch-off behaviour of a water filament using high-speed photography. When a droplet falls from an orifice under gravity, a column--like neck forms between the orifice and drop. The neck thins to a filament of some minimum diameter at which it breaks, and the droplet detaches. This process is highly surface-tension dependent, as a balance between inertial and surface tension forces causes the droplet to move and the neck to thin. Self-similar behaviour is exhibited at the filament [21] and the thinning close to pinch-off is universal [22], independent of the initial conditions of the experiment. For inviscid liquids, potential flow theory describes the thinning dynamics [23] and Keller [21] developed a universal scaling law that governs the thinning of the filament prior to breaking in the limit of low viscosity.

$$D_{min} = A(\gamma/\rho)^{1/3} \tau^{2/3} \qquad (8)$$

where $D_{min}$ is the minimum diameter of the filament indicated in Fig. 6(a), $A$ is a prefactor and $\rho$ and $\gamma$ are the density and surface tension of the fluid respectively. The time scale $\tau$, defined as $(t_0 - t)$, is the time to pinch--off at $t_0$. This expression is independent of gravity as the typical length scale in the necking process is much smaller than the capillary length $\kappa^{-1} = (\gamma/\rho g)^{1/2}$, a measure of the length below which gravity is negligible compared to surface tension forces. For water it is 2.72 mm.

This scaling law was predicted to hold for minimum filament radius values, $R_{min}$, much greater than the viscous length scale $l_v = \mu^2/\gamma\rho$, where $\mu$ is the dynamic viscosity[24]. As the filament thins past this point ($l_v = 14$ nm for water), the dynamics transition to the inertial-viscous regime where viscosity contributes [25]. There is considerable variation in the literature on values obtained for the universal prefactor in Eq. 8 which range from 0.2 to 1.46, from both numerical and experimental studies [23,26-29].

Hauner *et al* [17], showed that by measuring $D_{min}$ close to pinch-off, a *dynamic surface tension* for low viscosity liquids can be inferred using Eq. 8. The term is most used in the context of surfactant solutions. When a new surfactant interface is formed, its initial surface tension will decrease as surfactant molecules are transported to the interface until the equilibrium surface tension is reached [18]. In pinch-off of pure liquid, the dynamic surface tension inferred is the value at the liquid filament where the interface is in the process of forming on a millisecond timescale. For pure liquids, no discrepancy between the static and dynamic surface tension was expected. However, Hauner *et al*'s results [17] suggested a higher dynamic value for water, $\sim 90$ mNm$^{-1}$ compared to the static value of 72.8 mNm$^{-1}$, which they calculated with a prefactor of 0.9, determined by fitting experimental data for several liquids to Eq. 8 using the prefactor as a fit parameter. They suggested that the dielectric relaxation, typically at a scale of $\sim$ ps, may be much longer at the surface, with the time scale of relaxation during pinch-off being of order of $\sim 1$ ms.

It has since been shown that the prefactor in Eq. 8, that was assumed to be constant for all liquids, varies due to an influence of viscosity for minimum filament radii $D_{min} \gg l_v$ [19]. While the model may not then be used to infer an absolute quantitative value for the dynamic surface tension of the pinch-off system without further analysis, it allows us to investigate any relative differences in surface tension for pinch-off in the presence of a magnetic field. While previous studies have been carried out in electric fields [20] where no effect of field on the rate of pinch-off was found, ours is the first study in a magnetic field.

High-speed videos of the pinch-off of the fluid filaments were obtained using a Phantom v2010 camera with a Navitar 12× microscope lens as shown in Fig. 4. A plastic shield around the nozzle and magnet assembly mitigated effects from ambient air currents on the pinch-off process.

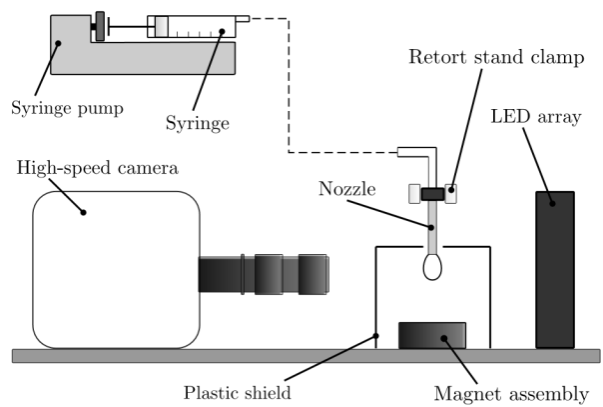
Figure 4. Experiment to record the droplet pinch-off.

Videos were taken at 89,000 fps, with an exposure time of 10 µs. Liquid was delivered to a quartz nozzle with outer diameter of 3.02 mm using a syringe pump at a constant rate of 100 µl/min to form a pendant droplet on the nozzle tip. Additional liquid was then added incrementally until the droplet reached the critical volume where it continued to deform in the absence of flow. This allows the droplets to detach under their own weight. The resolution of the pinch-off videos was 12 µm. Measurements were taken both in the absence of a magnetic field, and with two different magnet arrays. Fields 1 and 2, shown in Fig. 5 were produced using the 5-magnet array shown in Fig. 2(b), while fields 3 and 4 were produced with two parallel identical neodymium magnets with dimensions $25 \times 10 \times 50$ mm. For field 3, the field strength at the nozzle tip aligned with the top of the magnets is 137 mT. The field strength at the nozzle tip for field 4 is approximately 0.3 mT. The field strengths for fields 1 and 2 are given in Fig. 3(a).

The videos obtained were converted to images and the minimum diameter of the liquid filament was measured using a custom MATLAB script. Two different liquids were examined using this process – Millipore water and 0.11 M $DyCl_3$ solution. Images obtained for field direction 1 displayed an 8° tilt in the highly susceptible $DyCl_3$ filament due to the magnetic field gradient which can be seen in Fig. 6. Since the angle is small, the images were rotated for analysis so that the filament was vertical.

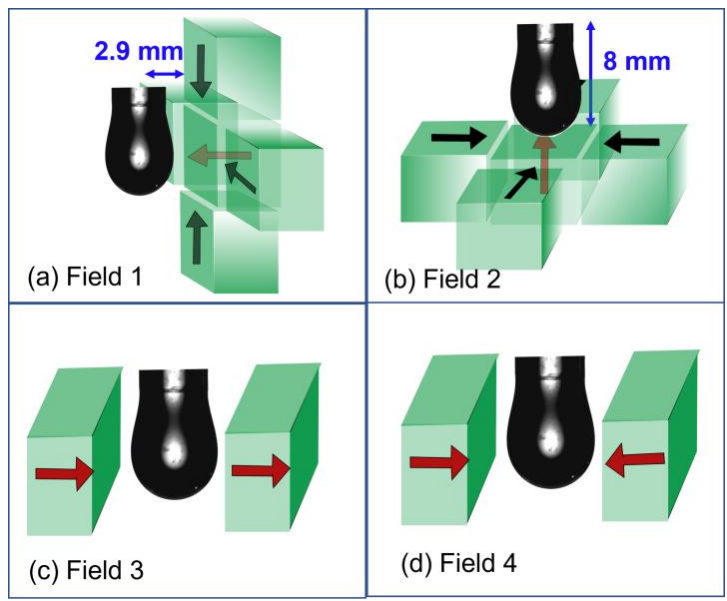

Figure 5. Field directions used in the dynamic experiment. In (a), the face of the 5-magnet array of Fig. 2(b) is aligned vertically with the centre opposite to the centre of a stable pendant drop on the nozzle, In (b) the nozzle was above the centre of the magnet array. In (c) and (d) two vertical magnets are aligned or opposed, respectively. The end of the nozzle was aligned with the top of the magnets

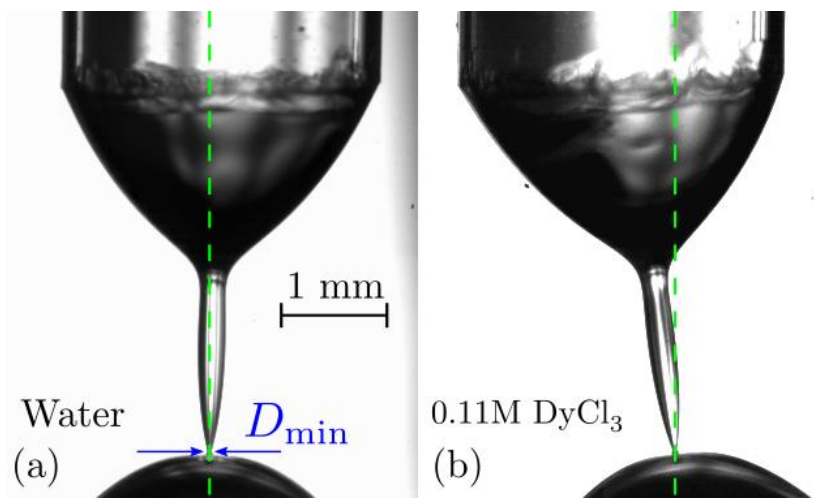

Figure 6. Fluid filament for (a) water and (b) 0.11M $DyCl_3$ 45 µs before pinch-off in field 1. The position of $D_{min}$ is indicated in (a).

# 3 Results and Discussion

## 3.1 Static measurements

The liquids used for these experiments were pure Millipore deionized water, and solutions of 99% pure $CuSO_4$, $MnSO_4$ or $DyCl_3$ in distilled water with the following molarities: Cu 0.1, 0.38, 0.42, 0.45, 0.47 and 0.56M; Mn 0.048 and 0.052 and Dy 0.01, 0.05, 0.1, 0.4 and 1M.

In order to determine the change of apparent surface tension $\gamma_{app}$ due to the magnetic field, measurements were made as follows: The liquid droplet was observed in the OCA 25 for an initial period of 60 seconds, then the magnet array was raised to the measurement position and the record continued for a further 60 seconds, before lowering the magnet and re-establishing the zero-field background for a final 60 seconds. In this way, any drift due to temperature change or evaporation could be corrected. This is important as it has been shown that added ions will change the rate of evaporation of a pendant drop [21]. Some representative results for water and 0.1M dysprosium solution are shown in Fig. 7, where we observed changes of apparent surface tension of 2.2 $mNm^{-1}$ for water and -5.1 $mNm^{-1}$ for dysprosium in the vertical gradient field of Fig. 3a) at position V1, and smaller values in the weaker vertical field gradient of Fig. 3c) at position H1. The body force acting on the droplet due to the magnetic field gradient changes its shape, which the instrument interprets as a change in surface tension, as the density and gravitational acceleration are fixed. In diamagnetic water, the effect is a compression of the droplet, upwards, and an increase in apparent surface tension, but for paramagnetic solutions, such as dysprosium chloride, the negative vertical magnetic field gradient results in a downward body force and that the droplet, corresponding to a decrease in apparent surface tension, as shown in Fig. 8.

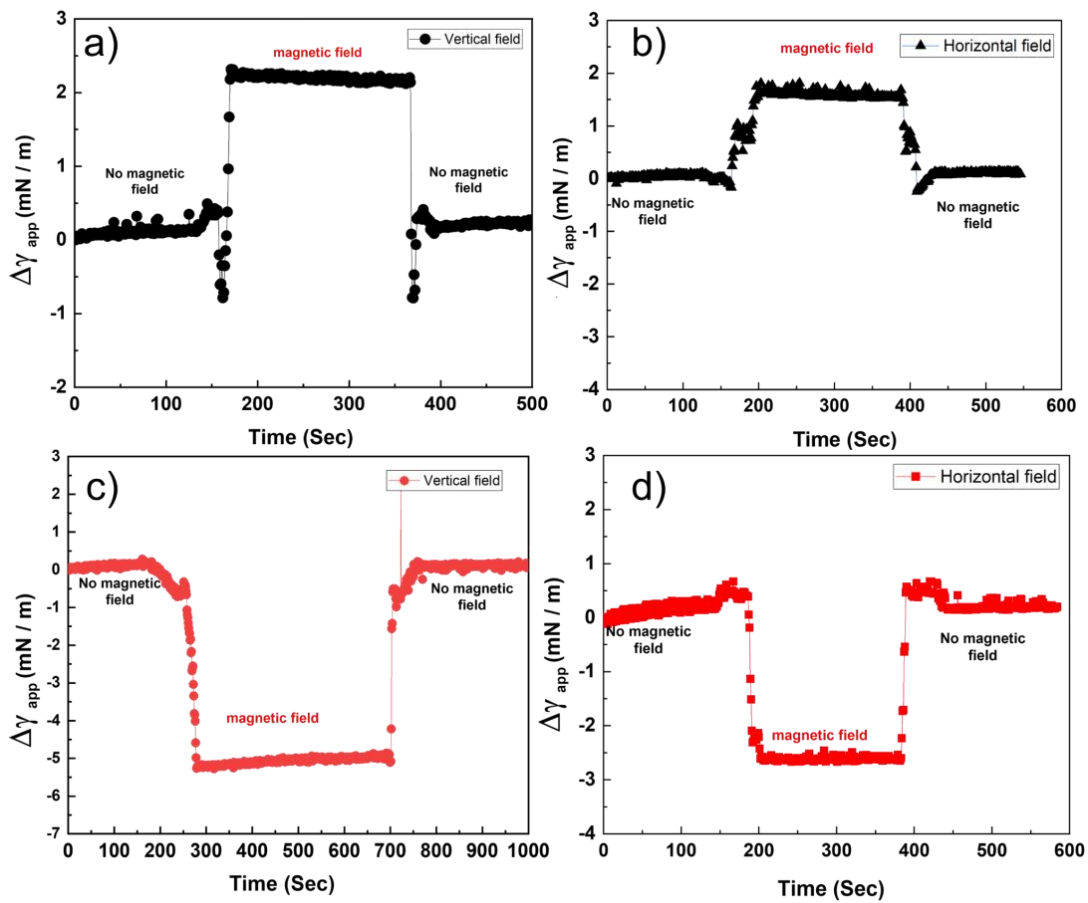

Figure 7. Changes of apparent surface tension change of water in a) a vertical gradient field, b) a horizontal gradient field and 0.1 M $DyCl_3$ in c) a vertical gradient field d) a horizontal gradient field in position 1.

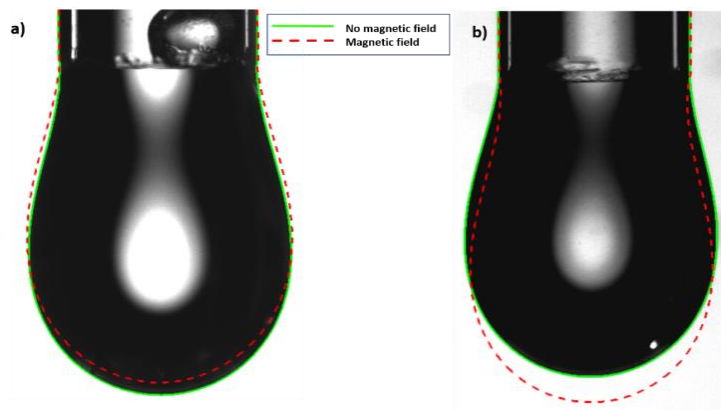

Figure 8: Images of a pendant droplet of a) water b) 0.1M $DyCl_3$ with and without magnetic field, with a negative vertical gradient. The water droplet is compressed, but the $DyCl_3$ droplet is extended.

$B\nabla_z B$ is 80 T²m⁻¹ for the 5-magnet vertical field structure and 14 T²m⁻¹ for the 3-magnet horizontal field structure in Fig. 3. The effective changes of density for water in the two cases would be -6% and -1%, while the corresponding changes for 0.1 M DyCl$_3$ are 39% and 7%, respectively. In Fig. 9a) we plot the measured apparent changes of surface tension $\Delta\gamma_{app}$, computed by the OCA 25, against the change of density $\Delta\rho_{eff}$, needed to fit the droplet shape at constant surface tension. An ordinary least-squares fit giveas a slope of the graph of -1.00 ± 0.02, as expected. In fig. 9b) we plot the change in gravitational force corresponding to the change of density needed to fit the droplet shape against the field gradient force obtained from the known values of $B\nabla_z B$ using Eq. 5 for force, both in kNm⁻³. Here the slope is 1.05 ± 0.09. This suggests that there is little change in surface tension due to the applied field.

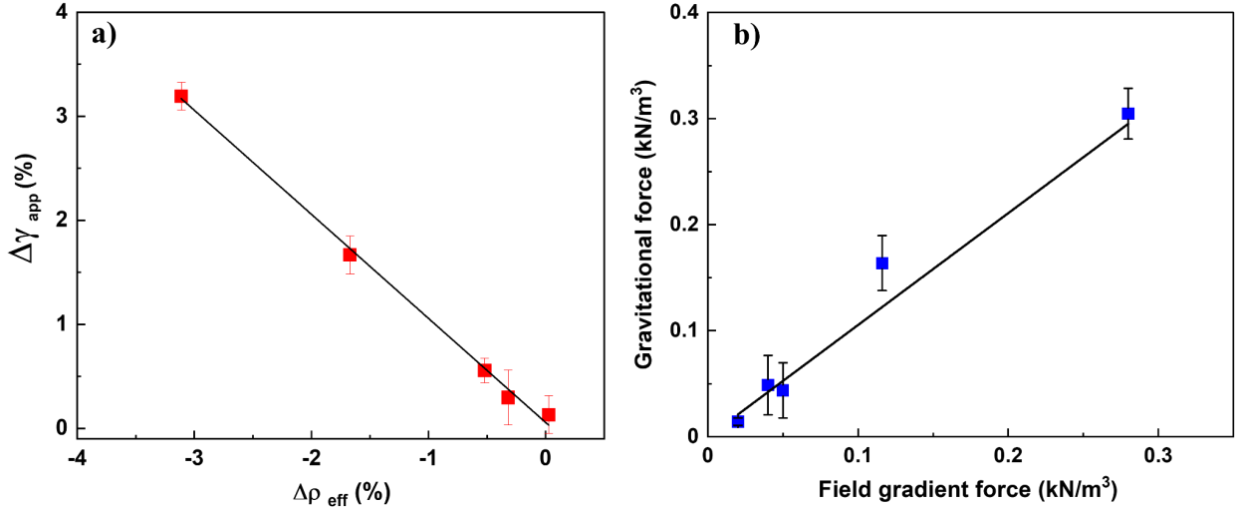

Figure 9. a) Plot of the measured change in apparent surface tension $\Delta\gamma_{app}$ versus $\Delta\rho_{eff}$ calculated from Eq. 5 versus for water and b) plot of gravitational force versus field gradient force for water.

For paramagnetic solutions, the magnetic field gradient force stretches the drop down towards the magnet and the extent of the stretching increases with solution concentration. The limiting factor to the stretching here is the shape of the pendant drop just before the pinch-off limit. There is a threshold extension for paramagnetic solution pendant drop, beyond which the drop detaches and sticks to the magnet, preventing further measurements. Thus the maximum stretching deformation is very much the same for 0.1 M Dy and 1 M Dy, as the critical nature of the necking process. All the measurements conducted use droplet shapes that are statically stable.

For water in a uniform field, the effect of Maxwell stress is negligible, and there is no field gradient so any effect must be attributed to a magnetically-induced change of the surface tension of water. We repeated the experiment 37 times, and the result was that the change of surface tension in the uniform 0.45 T field (Fig 3a) was $0.19 \pm 0.21$, mNm$^{-1}$, where the error is the standard deviation on the mean.

We use our zero susceptibility method to eliminate any magnetic body forces on the droplet, in order to look for any magnetic field effect on surface tension in magnetically undeformed droplets. Since any surface tension change is likely to be small, it was important to ensure that the solutions we are using really have zero susceptibility. We cannot rely on salt mass measurements before dissolution because the soluble salts are hygroscopic and readily absorb water. Furthermore, Eq. 7 is valid only in the dilute limit, when the density of water is independent of the dissolved ions. We measured the susceptibility of the solutions used directly in a SQUID magnetometer, in the range -5 T to 5 T. Some representative data for determining $x_0^{exp}$ for CuSO$_4$ solutions are shown in Fig. 10. The experimentally determined zero-susceptibility solutions are 472.7 mM, 51.3mM, and 17.18 mM for CuSO$_4$, MnSO$_4$ and DyCl$_3$, somewhat different from $x_0$ in Table 1. We then measured the change of surface tension induced by a magnetic field of 0.5 T as a function of concentration to the value of $\Delta\gamma$ when there is no magnetic body force. Some data are shown in Fig 10b.for Cu. where the change of surface tension for CuSO$_4$ in 0.5 T is ~ -0.4 mNm$^{-1}$. The corresponding results for MnSO$_4$ and DyCl$_3$ are -0.48 mNm$^{-1}$ and -0.30 mNm$^{-1}$.

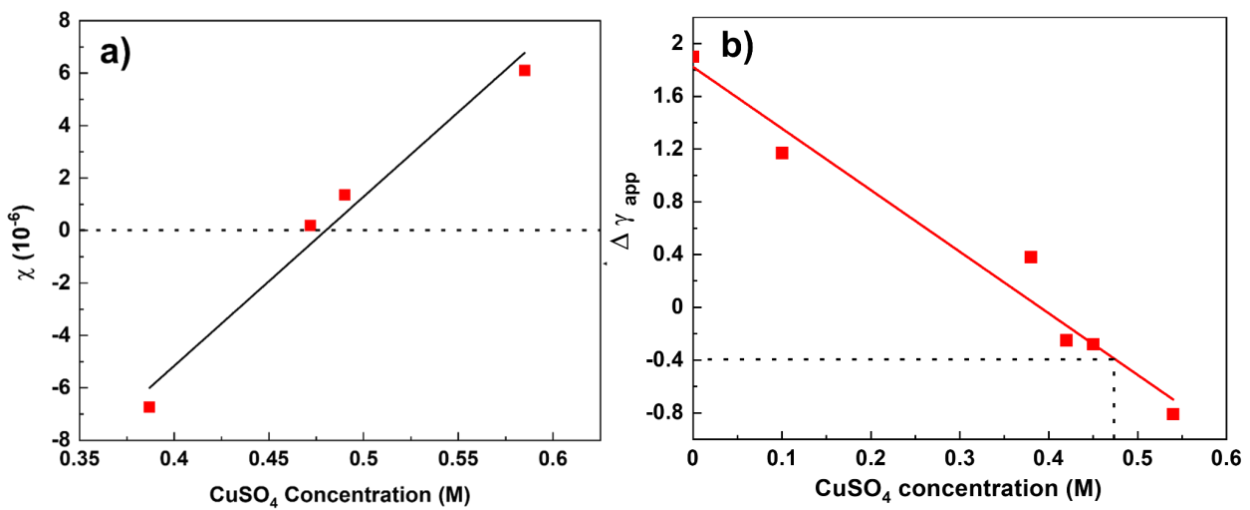

Figure 10. a) Zero susceptibility optimisation of different concentrations of CuSO$_4$, b) Variation of apparent surface tension as a function of the concentration of copper solutions

It is impossible to avoid subjecting the drop to a field gradient when the magnet is moved even when the horizontal homogenous field magnet is used.

Second order effects on droplet shape, going beyond the standard analysis in the introduction, can be treated by adding an effective gradient of the gravitational acceleration to the standard set of differential equations, Eq. 1 - 4. An example of this process is shown on Fig. 12. It can be seen that the quality of the shape parameter fit is improved by the introduction of this additional free parameter. Its reliable extraction, however, is hindered by the spatial resolution of the images (pixel density) and random pixel noise, which affect the extraction of the visible drop profile. The fitted values of the effective parameters for the fit of figure are: $\gamma_{eff} = 55.5$ mNm$^{-1}$, $\delta = -9.6$g m$^{-1}$. The high value of effective gradient of the gravitational acceleration is primarily due to the maximal field-gradient product of 80 T$^2$m$^{-1}$ and the substantial susceptibility of the Dy solution of 59 x10$^{-6}$. Magnetic field gradients thus provide a relatively convenient way to investigate the effects of inhomogeneous body forces (gravity) on liquid drops, without the complexity and cost of microgravity experimental environments, following the same underlying principles as magnetic levitation. As the solution used here is more than a factor of ten below the solubility saturation limit, there is potential to increase the magnitude of the observed effects by up to an order of magnitude, preserving the simplicity of the magnet arrays used.

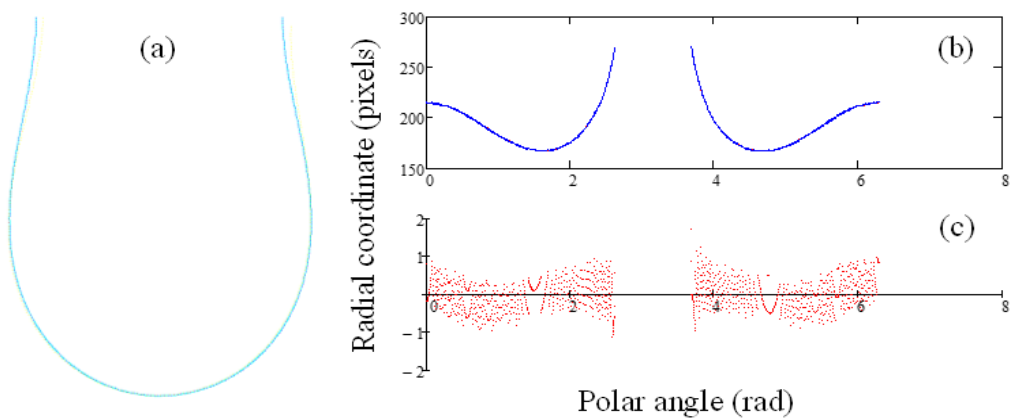

Figure 11. An example of the fitting process for a 0.1 M DyCl$_3$ solution, using the vertical gradient 5-magnet array. (a) a representation of the border of the drop`s projection – data in dots and continuous line for the fit; (b) the fitted data and fitting curve in polar coordinates; (c) fit residual in polar coordinates.

We end this section with a brief discussion of Maxwell stress[30,31]. Consider a spherical drop with susceptibility $\chi$ in a uniform horizontal magnetic field, as shown in Fig. 12. When $\chi \lesssim 10^{-3}$, demagnetizing effects are negligible. The drop has uniform magnetization $\mathbf{M} = \chi \mathbf{H}_0$, which can be represented by a surface magnetic charge density $\sigma_m = \mathbf{M} \cdot \mathbf{e}_n$, where $\mathbf{e}_n$ is the surface normal. This results in a tensile stress $\sigma$, as shown in the figure, and a spheroidal deformation of the drop, with half axes c and a, where $c = a(1+x)$. We compare the pressures at the points A and B, as shown. There is no magnetic pressure at A, since $\mathbf{M} \perp \mathbf{e}_n$, but at B $\mathbf{M} \parallel \mathbf{e}_n$, and $\sigma_m = \chi H_0$. The magnetic pressure is therefore $P_m = \mu_0 H_0 \sigma_m = \frac{1}{2} \chi \mu_0 H_0^2$. If $\mu_0 H_0 = 500$ mT and $\chi = 10 \times 10^{-6}$, $P_m = 1$ Pa. We equate this to the pressure difference between points A and B. Each is given by Eq. 1, where the two radii of curvature at point A are $c^2/a$ and a and at point B both are $a^2/c$. Setting $c = a(1 + x)$, we find $P_m$ to first order in z

$$P_B - P_A = 4x\gamma/a . \qquad (9)$$

Taking $a = 2$ mm, we find that the deformation of the drop is $x = 0.7$ %. The Maxwell stress would therefore produce an oblate compression of 0.6 % in the shape of a spherical drop of water and a 3.5% prolate extension shape of a spherical drop of 0.1M Dy solution. The effects in pendant droplets will be similar. These effects are small in relation to the droplet deformations illustrated in Fig. 8.

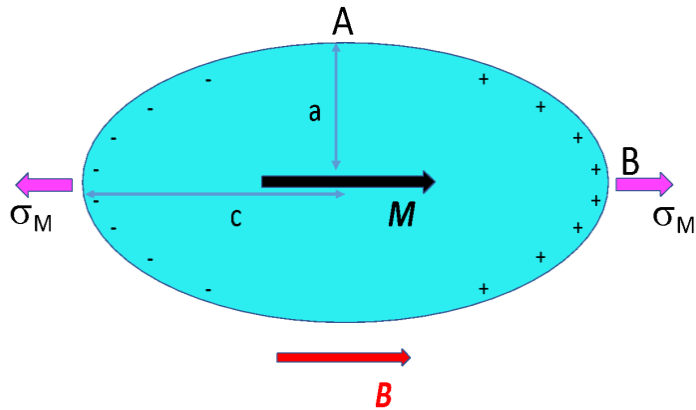

Figure 12 Prolate deformation of a spherical drop with a positive susceptibility due to Maxwell stress $\sigma_M$ due to the distribution of positive and negative magnetic charge induced by the magnetic field $\mathbf{B}$.

## 3.2 Dynamic measurements

To analyze the pinch-off data, $t_0$ was found for each of the 10 filaments measured per field. This is done by plotting $D_{min}$ as a function of $\tau$ for different values in the range where the data is linear and obeys the scaling law for $\tau < 0.4$ ms. Eq. 8 with a fixed slope of 2/3 is fitted to these data, and the value of $t_0$ that gives the fit is selected, illustrated in Fig. 13(a) for a single fluid filament of water in no field. $D_{min}^{2/3}$ can then then be plotted against $\tau$ and the slope of a linear fit is used to calculate the dynamic surface tension using a rearrangement of Eq. 8,

$$\text{slope} = A^{3/2} (\gamma_{DST}/\rho)^{1/2} \qquad (10)$$

where $\gamma_{DST}$ is the apparent dynamic surface tension, given the density of the fluid and some prefactor value $A$, which we know from Deblais *et al* [19] is not constant. Here, we are interested in the relative changes of the pinch-off dynamics in the presence of a magnetic field, rather than the calculated dynamic surface tension values. Assuming a constant prefactor of 0.9 as determined by Hauner [17], a dynamic surface tension of $97.3 \pm 0.03$ mN/m was obtained for water in no field. To compare measurements for different fields, the $D_{min}$ data for all droplets in each field configuration was averaged using built-in statistical analysis in Origin.

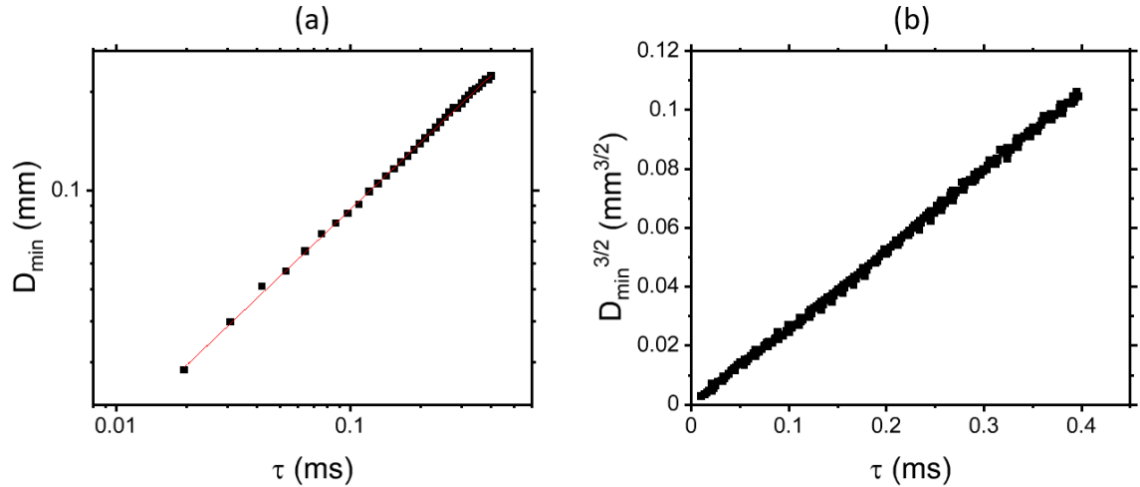

Figure 13. (a) Log-log fit of Eq. 8 for a single fluid filament to find $t_0$. (b) linear fit to determine the dynamic surface tension of water.

Figures 14(a) and 15(a) show the change in minimum filament radius $D_{min}$ with the dimensionless time $\tau^*$ given by $\tau^* = \tau/t_c$, where the capillary time $t_c = (\rho D_0^3/\gamma)^{1/2}$, $D_0$ is the nozzle diameter and $\gamma$ the static surface tension. Fig. 14(a) shows no variation in the averaged $D_{min}$ for water in each field configuration. The data all lie upon the same curve, indicating that the dynamic surface tension of water is unchanged in these different fields. The error is denoted by the shaded regions (standard deviation of averaged data, ranging from $3\times10^{-4}$ to $2\times10^{-3}$ mm). In Fig. 15(a) we see no significant difference between in-field and no-field data across all magnet configurations, this time for 0.11 M $DyCl_3$.

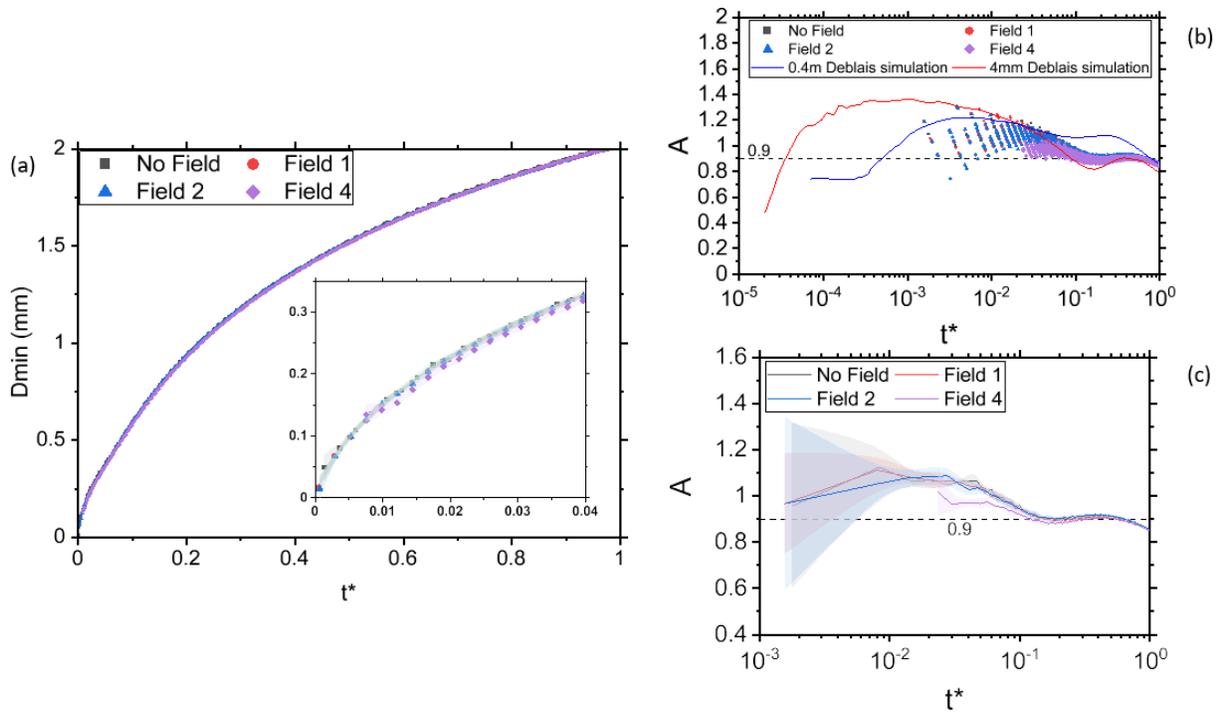

Figure 14. Pinch-off results for water. (a) Averaged data with data close to pinch-off in inset, (b) calculated prefactor as a function of dimensionless time t* (c) calculated prefactor as a function t* with averaged data.

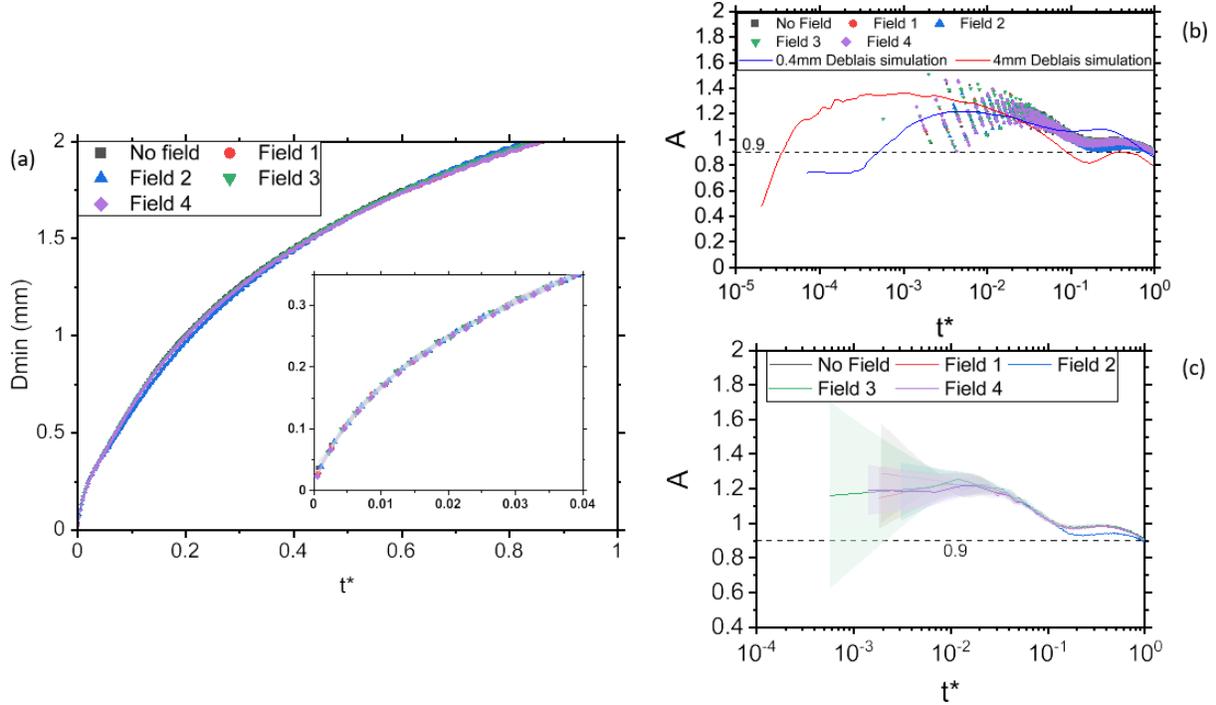

Figure 15. Pinch-off results for $DyCl_3$. (a) Averaged data with data close to pinch-off in inset, (b) calculated prefactor as a function of dimensionless time $t^*$ (c) calculated prefactor as a function $t^*$ with averaged data.

Analysis was also carried out to investigate how the prefactor $A$, changes over time, as shown by Deblais *et al* [19]. The prefactor was calculated by restating Eq. 8 in terms of the dimensionless time $t^*$, to obtain $(D_{min}/D_0)^{3/2} = A^{3/2} t^*$. $A$ is then calculated over the range of measured filament diameters, using literature values of static surface tension and density to calculate $t^*$. The results of the calculations are plotted in Figs 14 and 15 (b) and (c) for water and 0.11 M $DyCl_3$, respectively. Calculated prefactors over time for each individual filament are shown in panel (c) of each figure, with simulations from Deblais *et al* overplotted for reference. Note that these simulations are for different nozzle diameters than those used in this study and are not expected to agree perfectly with our results. Prefactor values for 10 different filament measurements are included for each field. Panel (c) of each figure shows an average prefactor value for each field over time with standard deviation indicated by the shaded regions. The spread in data points close to pinch-off seen in all plots is due to the increased error in $D_{min}$ measurements as the detected diameter approaches the resolution limit of 12 µm.

In Fig. 14(b) and (c), we can see the instantaneous prefactor values for each field follow the same trend. The data follows a similar trend to that predicted by the simulations, showing that the prefactor is not constant over the pinch--off process. However, for t* ≳ 0.1 it is close to 0.9 as measured by Hauner *et al* [17]. The fluctuating prefactor is also seen in Fig. 15(b) and 15(c) for 0.11 M $DyCl_3$. Values for the different field configurations follow the same trends, with the exception of field 2, at times far from pinch-off. Here the prefactor values lie below that of other fields, most apparent in Fig. 15(c).

Our plots for minimum filament diameter during the pinch-off process show no variation in behaviour for any of our four magnetic field configurations, which include both gradient and quasi-uniform fields. Our data follow the same trend for water and 0.11 M $DyCl_3$, indicating no change in dynamic surface tension for the paramagnetic solution in the presence of a magnetic field. The value reported here for the dynamic surface tension of water under no magnetic field, while agreeing that the dynamic surface tension of water is higher than that of the static surface tension, differs from that previously reported in the literature. Our result is 8% higher than the value measured by Hauner et al. [17] using the prefactor 0.9 they found experimentally. However, they assumed a constant prefactor. Our measured instantaneous prefactor results across both fluids show similar variations to those found by Deblais *et al* [19].

Our important finding is that the known magnetic body forces acting on the droplet do not influence the surface tension driven nature of the pinch-off process. In a gradient field, our fluid filament is acted upon by the Kelvin force, modifying the effective force of gravity on the filament. Pinch-off dynamics are not affected by gravity, as is evident by the absence of a term in the scaling law, Eq. 8. This is because pinch-off occurs at filament diameters far below the capillary length, 2.72 mm for water, 2.38 mm for 0.11 M $DyCl_3$. A small percentage change in the effective gravitational force due to the presence of magnetic body forces would barely change the capillary number, and so we would not expect to see an effect of these forces during pinch-off. The effective gravity, $g_{eff}$, can be calculated from the 8° angle of filament tilt shown in Fig. 8 for $DyCl_3$ in field 2 due to the high paramagnetic susceptibility of the filament. We obtain a capillary length of 2.36 mm and pinch-off remains unaffected by body forces due to the field. We do however see some deviation of the prefactor variation in field 2 for the $DyCl_3$ filament at times far from pinch-off figure 15 (c). The scaling law does not apply in this regime (1 mm < $D_{min}$ <1.5 mm) and body forces can no longer be neglected, resulting in deviations from the thinning dynamics in the absence of a field. We do not see this effect for measurements with water in field 2, as the change in the effective gravity due to the additional body forces due to the field is negligible due to the lower magnetic susceptibility of water.

## 4 Conclusions

Magnetic field effects on surface tension can be evaluated from droplet shape in either of two ways — in a uniform magnetic field, or in aqueous solutions of paramagnetic ions where the Curie-law paramagnetism cancels the diamagnetism of water to give zero net magnetic susceptibility, and therefore no magnetic body force. The effects are small; the first method applied to water gives changes that are barely significant, $0.42 \pm 0.47$ mNm$^{-1}$T$^{-1}$. The second method, applied to zero-susceptibility solutions of $Cu^{2+}$, $Mn^{2+}$ and $Dy^{3+}$ gives negative values in the range -0.77 to -1.13 mNm$^{-1}$T$^{-1}$. These measurements should be extended to much larger, uniform magnetic fields.

The shapes of pendant droplets of water and paramagnetic solutions are mainly deformed by the magnetic field gradients produced by permanent magnet arrays. Uniform vertical field gradients over the droplet volume are equivalent to changes of effective liquid density. The corresponding changes of droplet shape are interpreted by the droplet analyser as changes of 'apparent surface tension'. Nonuniform magnetic field gradients, however, produce droplet deformations that cannot be interpreted using the standard fitting algorithm. The droplet shapes can be fitted by adding a term with a gradient of gravitational acceleration into the differential equations. These second-order effects are dominant for strongly-paramagnetic solutions and have been calculated for 0.1 M $DyCl_3$. An 'effective surface tension' of 55.5 mNm$^{-1}$ has been measured in 0.1M $DyCl_3$ in an 'effective gravity gradient' of 0.1 ms$^{-2}$/mm. Such analysis of the non-uniform Kelvin force over a pendant drop requires high-resolution images of the droplet shape. The standard fitting algorithms are inapplicable for such solutions in a nonuniform magnetic field. Maxwell stress makes a minor contribution to the deformation.

Our method of measuring the static droplet deformation, which images the droplet in zero field before and after imaging it in the presence of the field, involves moving the permanent magnet array to and from the droplet and exposing it to a variable, transient field gradient. The method provides the zero-field baseline and allows the correction of small drifts in the course of a measurement, but there is an upper limit to the susceptibility that can be measured in a given magnet array, beyond which the droplet is pulled off when the magnets are first moved towards it. Measurements of changes in 'apparent surface tension' are limited to about 6 mNm$^{-1}$

No influence of magnetic field on the dynamic surface tension of water or Dy$^{3+}$ solution was detected in any of four magnetic configurations used within the error of the experiments. The pinch-off is unaffected by the field-gradient forces, which are equivalent to a change of gravity, because the filament diameter in the scaling regime is much less than the capillary length. The negative changes in static surface tension observed for paramagnetic solution of about -1 mNm$^{-1}$T$^{-1}$ are within the error of the dynamic surface tension measurements.

**Acknowledgements** Support from the European Commission from contract No 766007 for the 'Magnetism and Microfluidics' Marie Curie International Training Network is acknowledged, as well as support from Science Foundation Ireland contracts 12/RC/2278 AMBER, 16/RI/3403, 16/A/4534, and 17/CDA/4704. We are very grateful to Dr. M. Venkatesan for his help with this work, and to Peter Dunne for help with field profiling of the magnetic arrays.

**Author Contributions.** MEM and JMDC formulated the project and supervised the work. SP and JAQ carried out the experiments and analysed the data. PS analysed the droplet shape in a nonuniform field gradient. All authors discussed the results. SP, JAQ and JMDC wrote the paper.